\documentclass{moriond}
\usepackage[utf8]{inputenc}
\usepackage{amsfonts}
\usepackage{amsmath}
\usepackage{amssymb}

\newcommand{\Photo}{}

\begin{document}
\setlength{\parindent}{0pt}
\phantom{------------------------------------------------------------------------------------------------------------} DO-TH 22/15
\vspace*{4cm}

\title{The role of dineutrino modes in the search for new physics}

\author{Hector Gisbert}

\address{Fakult\"at Physik, TU Dortmund, Otto-Hahn-Str.\,4, D-44221 Dortmund, Germany}

\maketitle\abstracts{Dineutrino modes offer promising searches for new physics. The potential aspects of these modes are reviewed in detail. Performing a proper combination of them, novel tests of the SM symmetries are derived. Different phenomenological applications are worked out, including charm, beauty and kaons, which result in novel tests of lepton universality and charged lepton flavour conservation with flavour-summed dineutrino observables, in addition to improved bounds on $\tau\,\ell$ couplings with $\ell=e,\mu,\tau$.}

\section{Introduction}

Flavour-changing neutral currents (FCNCs) of $q^\alpha$ and $q^\beta$ quarks driven by $|\Delta \,q^\alpha|=|\Delta\, q^\beta|=1$ processes represent excellent probes of physics beyond the Standard Model (SM). In the SM, FCNCs are forbidden at tree level and are strongly suppressed in loop corrections by the Glashow–Iliopoulos–Maiani (GIM) mechanism and Cabibbo-Kobayashi-Maskawa (CKM) hierarchies. This strong suppression, not necessarily present in SM extensions, can result in large experimental deviations from the SM predictions alluding to a breakdown of SM symmetries. Further potential SM tests emerge if leptons are involved, such $q_\alpha\, q_\beta\,{\ell^\prime}^+\,{\ell}^-$ and $q_\alpha\, q_\beta\,\bar\nu_{\ell^\prime}\,\nu_{\ell}$ interactions. We exploit the $SU(2)_L$-invariance between left-handed charged lepton and neutrino couplings. This connection can be established using the Standard Model Effective Field Theory (SMEFT) framework~\cite{Grzadkowski:2010es} and allows to perform complementary experimental tests of Lepton Universality (LU) and charged Lepton Flavour Conservation (cLFC) with flavour-summed dineutrino observables~\cite{Bause:2020auq}.

These proceedings are organized as follows: In Section~\ref{sec:theory}, we highlight the essential features of dineutrino modes in the SM and beyond. Some of these aspects are rather simple and well-known in the particle physics community, however, the proper combination of them leads to novel phenomenological implications, as shown in Section~\ref{sec:pheno}. We conclude in Section~\ref{sec:con}. The results are based on Refs.~\cite{Bause:2020auq,Bause:2020xzj,Bause:2021cna}, we refer there for further details and other interesting results.

\section{The fingerprints of dineutrino modes}\label{sec:theory}

Assuming the same particle content as the SM below the electroweak scale, $|\Delta q^\alpha| = |\Delta q^\beta| = 1$ dineutrino transitions can be described by the following effective low-energy Hamiltonian~\footnote{Wilson coefficients in calligraphic style denote those for mass eigenstates.}
\begin{equation} \label{eq:Heff_noright}
	\mathcal H_\text{eff}^{\nu_\ell\bar{\nu}_{\ell^\prime}} = -\frac{4\,G_\text F}{\sqrt 2}\frac{\alpha_e}{4\pi} \left( \mathcal{C}_L^{N_{\alpha \beta}\ell\ell^\prime} Q_L^{N_{\alpha \beta}\ell\ell^\prime} + \mathcal{ C}_R^{N_{\alpha \beta}\ell\ell^\prime}Q_R^{N_{\alpha \beta}\ell\ell^\prime} \right) + \,\text{h.c.}\,,
\end{equation}
with only two four-fermion operators
\begin{align}\label{eq:opsneu} 
Q_{L(R)}^{N_{\alpha \beta}\ell\ell^\prime}&=  (\bar q_{L(R)}^\beta\, \gamma_\mu \,q_{L(R)}^\alpha)  \,  (\bar \nu_{L}^{\ell^\prime} \,\gamma^\mu\, \nu_{L}^\ell) \, .
\end{align}
Here, $N=U$ ($N=D$) represents the up-quark (down-quark) sector, while the indices $\alpha,\beta$ and $\ell,\ell^\prime$ denote the quark and the neutrino flavours, respectively. For charged dilepton transitions (where further dimension six operators exist), we use $\mathcal{K}_{L,R}^{N_{\alpha\beta}\ell\ell^\prime}$ and $O_{L,R}^{N_{\alpha\beta}\ell\ell^\prime}$ for Wilson coefficients and operators, respectively. From now on (and if the opposite is not stated), we use the following shortened notation, $\mathcal{C}_{L,R}^{bs\ell\ell^\prime}=\mathcal{C}_{L,R}^{D_{23}\ell\ell^\prime} $, $Q_{L,R}^{bs\ell\ell^\prime}=Q_{L,R}^{D_{23}\ell\ell^\prime} $, and similar for other flavour combinations.

\subsection{Strong GIM suppression and lepton universality in the Standard Model}
Due to the $V-A$ structure of the weak interations, only the operator $Q_L^{\alpha\beta\ell\ell^\prime}$ receives non-zero contributions in the SM, and $\mathcal{C}_{L,R,\,\text{SM}}^{\alpha\beta\ell\ell^\prime}$ become~\footnote{The convention $(1,2,3)=(u,c,t)$ and $(d,s,b)$ is adopted.} 
\begin{align}\label{eq:CL_SM}
    \mathcal{C}_{L,\,\text{SM}}^{\alpha\beta\ell\ell^\prime}\,\approx\,\delta_{\ell\ell^\prime}\left(  \lambda_3^{(\alpha\beta)}\,f(x_3)\,+\lambda_2^{(\alpha\beta)}\,f(x_2)\right),\quad \mathcal{C}_{R,\,\text{SM}}^{\alpha\beta\ell\ell^\prime}\,=\,0~,
\end{align}
after applying the unitarity of the CKM matrix. The parameter $\lambda_i^{(\alpha\beta)}=V_{\alpha i}\,V_{\beta i}^*$ encodes the dependence on the CKM matrix elements, and the function $f(x_i)\sim x_i/(2\,\text{sin}^2\theta_W)$ with $x_i=m_{i}^2/M_W^2$ parameterizes the quantum effects from the $Z$-penguin diagram.

Using Eq.~\eqref{eq:CL_SM} together with the branching ratio of a hadron $h$ (with quark content $q^\alpha$, mass $m_h$ and lifetime $\tau_h$) decaying into a final hadronic state $F$ (with content $q^\beta$ and mass $m_F\ll m_h$) and two neutrinos, given by $\mathcal{B}(h\to F\,\nu\,\bar{\nu})\,\approx\,(\tau_{h}\,G_F^2\,\alpha_e^2\, m_{h}^3\,|\mathcal{C}_L^{\alpha\beta\ell\ell}|^2)/(16\,(2\,\pi)^5)$, we obtain the following estimations for different quark transitions
\begin{align}\label{eq:naive_estimation_BR}
    \mathcal{B}(b\to s \,\nu\,\bar{\nu})\sim  10^{-6},\,\,\mathcal{B}(b\to d\, \nu\,\bar{\nu})\sim  10^{-7},\,\,\mathcal{B}(s\to d\, \nu\,\bar{\nu})\sim  10^{-8},\,\,\mathcal{B}(c\to u \,\nu\,\bar{\nu})\sim  10^{-19}~.
\end{align}
Eq.~\eqref{eq:naive_estimation_BR}, although being very naïve, exhibits the strong GIM suppression of dineutrino modes in the SM, which is particularly efficient for $c\to u\,\nu\bar{\nu}$ transitions. 

In the SM, the leptonic interactions of the three families of fermions are identical, except for the different masses of the constituent particles. We can observe this pattern in Eq.~\eqref{eq:CL_SM}, where the $Z$ boson couples in the same manner to the three lepton generations (see the Kronecker delta $\delta_{\ell\ell^\prime}$). This peculiar aspect of the SM is known as LU, and it is broken at the Lagrangian level by corrections proportional to the lepton masses ({\it i.e.} the Yukawa interaction between the Higgs field and the lepton fields), or directly in the observables by kinematic differences due to the masses. To test lepton flavour violation, it is crucial to design observables where such effects are under control. Thanks to the strong suppression of the neutrino masses, $q^\alpha\to q^\beta\, \nu\,\bar\nu$ transitions provide very clean observables, in contrast to $q^\alpha\to q^\beta \,\ell^-\ell^+$ decays where these effects can be sizeable (specially if $\tau$ leptons are involved). 

\subsection{Lepton flavour inclusiveness}\label{subsec:inclusiveness}

Since the neutrino flavours are experimentally untagged, any dineutrino observable $\mathcal{O}(\nu_\ell\,\bar{\nu}_{\ell^\prime})$ requires an incoherent sum over lepton flavours, that is
\begin{align}\label{eq:Obs}
    \mathcal{O}(\nu\,\bar{\nu})\,=\,\sum_{\ell\ell^\prime}\mathcal{O}(\nu_\ell\,\bar{\nu}_{\ell^\prime})\,=\,p\,\mathcal{O}_{\text{max}}(\nu_\ell\,\bar{\nu}_{\ell^\prime})~,\quad p\,=\, \frac{\sum_{\ell\ell^\prime}\mathcal{O}(\nu_\ell\,\bar{\nu}_{\ell^\prime})}{\mathcal{O}_{\text{max}}(\nu_\ell\,\bar{\nu}_{\ell^\prime})}\,\leq\, 9~.
\end{align}
Here, $\mathcal{O}_{\text{max}}(\nu_\ell\,\bar{\nu}_{\ell^\prime})$ represents the largest contribution among all $\ell,\ell^\prime$ configurations. Interestingly, if LU is preserved, $\mathcal{O}(\nu_\ell\,\bar{\nu}_{\ell^\prime})_{\text{LU}}=\delta_{\ell\ell^\prime}\, \mathcal{O}_{\text{max}}(\nu_\ell\,\bar{\nu}_{\ell^\prime})$, then the sensitivity of $\mathcal{O}(\nu\,\bar{\nu})$ gets enhanced by a factor $p=3$. In contrast if new physics (NP) allows for non-zero values of $\mathcal{O}(\nu_\ell\,\bar{\nu}_{\ell^\prime})$ with $\ell\neq \ell^\prime$, $p$ can get further enhanced (up to $p=9$ if $\mathcal{O}(\nu_\ell\,\bar{\nu}_{\ell^\prime})=\mathcal{O}_{\text{max}}(\nu_\ell\,\bar{\nu}_{\ell^\prime})$ for any $\ell,\ell^\prime$ combination). Such aspect is absent in charged dilepton observables $\mathcal{O}(\ell^+{\ell^\prime}^-)$
where the lepton flavour is experimentally tagged.

\subsection{Correlations between dineutrino observables and EFT tests}\label{subsec:EFTtest}

Thanks to Lorentz invariance and parity conservation in the strong interaction, any dineutrino observable $\mathcal{O}(\nu\bar\nu)$ depends on at most two combinations of Wilson coefficients
\begin{align}\label{eq:xpm}
    x^\pm_{\alpha\beta}\,=\,\sum_{\ell\ell^\prime}\left|\mathcal{C}_L^{\alpha \beta\ell\ell^\prime}\pm\mathcal{C}_R^{\alpha \beta\ell\ell^\prime}\right|^2~,
\end{align}
where the different signs between $x^\pm_{\alpha\beta}$ reflect the excellent complementarity between both parameters, $x_{\alpha\beta}=(x_{\alpha\beta}^++x_{\alpha\beta}^-)/2=\sum_{\ell\ell^\prime}|\mathcal{C}_L^{\alpha \beta\ell\ell^\prime}|^2+\sum_{\ell\ell^\prime}|\mathcal{C}_R^{\alpha \beta\ell\ell^\prime}|^2\leq 2\, x^\pm_{\alpha\beta}$. In principle, only two experimental measurements are required to fully describe the short-distance dynamics of dineutrino modes. Assuming we have access to three experimental measurements, $\mathcal{O}_{1,2,3}(\nu\bar{\nu})$,  and considering that they depend linearly on $x^\pm_{\alpha\beta}$ (like in the branching ratios)
\begin{align}\label{eq:observable}
\mathcal{O}_i(\nu\bar{\nu})\,&=\,A_+^i\,x^+_{\alpha\beta} +A_-^i\,\,x^-_{\alpha\beta}~,\quad i=1,2,3~,
\end{align}
where the parameters $A_\pm^i$ encode the long-distance dynamics and the kinematics of the $q^\alpha\to q^\beta$ transition involved in $\mathcal{O}_i(\nu\bar{\nu})$. The three observables lead to an over-constrained system
with more observables than unknown parameters, or in other words to a strong  correlations between them~\footnote{Note that Eq.~\eqref{eq:deltaOcorr} is the cross product of the vectors $r(A_\pm^i,A_\pm^j)$ and $\mathcal{O}_k(\nu\bar{\nu})$. If these vectors are not linearly independent, or if either one has zero length, then their cross product is zero. Geometrically, $\delta\mathcal{O}_{\text{corr}}$ can be seen as the area of the triangle form by the three observables $\mathcal{O}_{1,2,3}(\nu\bar{\nu})$ in the plane with $x^{+}_{\alpha\beta}$-- and $x^{-}_{\alpha\beta}$--axes. In our EFT framework~\eqref{eq:Heff_noright} all lines cross in the same $(x^{+}_{\alpha\beta},x^{-}_{\alpha\beta})$ point, and therefore the area is zero.}
\begin{align}\label{eq:deltaOcorr}
   \delta\mathcal{O}_{\text{corr}}\,=\,
   \frac{1}{2}\sum_{i,j,k}\epsilon_{ijk}\, r(A_\pm^i,A_\pm^j)\,\mathcal{O}_k(\nu\bar{\nu})=0~,
\end{align}
where $r(A_\pm^i,A_\pm^j)=A_+^i\,A_-^j\,-A_-^i\,A_+^j$. Eq.~\eqref{eq:deltaOcorr} allows us to test our EFT assumptions, since any misalignment, $\delta\mathcal{O}_{\text{corr}}\,\neq\,0$, would clearly hint to the presence of additional Wilson coefficients, not considered in Eq.~\eqref{eq:Heff_noright}, {\it i.e.} scalar and tensor operators built with light right-handed neutrinos. Another possibility can be an observable where the contributions~\eqref{eq:opsneu} are highly suppressed by the mass of the light neutrinos. A clear example is the branching ratio of a meson $M^0$ decaying into two neutrinos, $\mathcal{B}(M^0\to\nu\bar{\nu})$ , where the contributions~\eqref{eq:opsneu} suffer a strong chiral suppression of two powers of the neutrino masses. Notice that, in the EFT context of Eq.~\eqref{eq:Heff_noright}, we can use Eq.~\eqref{eq:deltaOcorr} to indirectly improve the limits of some observables if others have better limits.

\subsection{$SU(2)_L$--link between dineutrino and charged dilepton couplings}

The lowest order Lagrangian accounting for semileptonic (axial-)vector four-fermion operators in SMEFT reads~\cite{Grzadkowski:2010es}
\begin{align}\label{eq:ops} 
{\mathcal{L}}_{\text{eff}} & \supset \frac{C^{(1)}_{\ell q}}{v^2}\, \bar Q \gamma_\mu Q \,\bar L \gamma^\mu L +\frac{C^{(3)}_{\ell q}}{v^2}\, 
\bar Q \gamma_\mu  \tau^a Q \,\bar L \gamma^\mu \tau^a L   +\frac{C_{\ell u}}{v^2} \, \bar U \gamma_\mu U \,\bar L \gamma^\mu L +
\frac{C_{\ell d }}{v^2} \,\bar D \gamma_\mu D \,\bar L \gamma^\mu L \,. 
\end{align}
Expanding~\eqref{eq:ops} into $SU(2)_L$-components, one obtains the couplings to dineutrinos $(C_{L,R}^N)$ and charged dileptons $(K_{L,R}^N)$, 
\begin{align}\label{eq:links} 
\begin{split}
C_L^{U\,(D)}&=K_L^{D\,(U)}={\frac{2\pi}{\alpha}}\left(C^{(1)}_{\ell q} +\,(-)\, C^{(3)}_{\ell q}\right)  \,, \, C_R^{U\,(D)}=K_R^{U\,(D)}={\frac{2\pi}{\alpha}}C_{\ell u(d)}  \,, 
\end{split}
\end{align}
Interestingly, $C_R^N=K_R^N$ holds model-independently. However due to the different relative signs of $C^{(1)}_{\ell q}$ and $C^{(3)}_{\ell q}$, the Wilson coefficient $C_L^N$ is not fixed by $K_L^N$ in general. Writting Eqs.~\eqref{eq:links} in the mass basis ($\mathcal{C}_{L}^{N} = W^\dagger  \,\mathcal{K}_{L}^{M}\, W+ {\mathcal{O}}(\lambda)$, $\mathcal{C}_{R}^{N}=W^\dagger \,\mathcal{K}_{R}^{N}\, W$ where $W$ is the Pontecorvo-Maki-Nakagawa-Sakata (PMNS) matrix and $\lambda \sim 0.2$ the Wolfenstein parameter) and inserting them into $x_{\alpha\beta}$, one obtains the following identity between charged lepton couplings $\mathcal{K}_{L,R}$ and neutrino ones $\mathcal{C}_{L,R}$~\cite{Bause:2020auq}
\begin{align}
   &x_{\alpha\beta}\,=\,\sum_{\ell,\ell^\prime}  \big\vert\mathcal{C}_L^{{N} \ell\ell^\prime}\big\vert^2+\sum_{\ell,\ell^\prime}\big\vert\mathcal{C}_R^{{N} \ell\ell^\prime}\big\vert^2=\sum_{\ell,\ell^\prime} \big\vert\mathcal{K}_L^{{M} \ell\ell^\prime}\big\vert^2+\sum_{\ell,\ell^\prime}\big\vert\mathcal{K}_R^{{N} \ell\ell^\prime}\big\vert^2 + {\mathcal{O}}(\lambda)\,=\,\sum_{\ell,\ell^\prime}r_{\ell\ell^\prime} ~,\label{eq:super} 
\end{align}
where the trace identity $\text{Tr}(X\,X^\dagger)=\sum_{\ell,\ell^\prime}|X_{\ell\ell^\prime}|^2$ and the unitarity of the PMNS matrix have been used. Here, $r_{\ell\ell^\prime}=\vert\mathcal{K}_L^{{M} \ell\ell^\prime}\vert^2+\vert\mathcal{K}_R^{{N} \ell\ell^\prime}\vert^2 + {\mathcal{O}}(\lambda)$; we use $N,M=U,D$ ($N,M=D,U$) when the link is exploited for neutrino couplings in the up-quark sector (in the down-quark sector). Eq.~\eqref{eq:super} allows the prediction of dineutrino rates for different leptonic flavor structures $\mathcal{K}_{L,R}^{N \, \ell\ell^\prime}$,
\begin{itemize}
\item[{\it i)}] $\mathcal{K}_{L,R}^{N \, \ell\ell^\prime} \propto \delta_{\ell\ell^\prime}$,  \textit{i.e.} \emph{lepton-universality} (LU),
\item[{\it ii)}] $\mathcal{K}_{L,R}^{N \, \ell\ell^\prime}$ diagonal, \textit{i.e.}  \emph{charged lepton flavor conservation} (cLFC),
\item[{\it iii)}] $\mathcal{K}_{L,R}^{N \, \ell\ell^\prime}$ arbitrary,
\end{itemize}
which can be probed with lepton-specific measurements. 

\section{Phenomenological applications}\label{sec:pheno}

In the following, we exploit the exceptional properties of dineutrino modes (presented in Section~\ref{sec:theory}) for different phenomenological applications. Consequently, we obtain novel lepton flavour violation tests and currently the strongest constraints for some charged dilepton couplings.

\subsection{Predictions for charm}

Due to the efficient GIM-suppression, seen naïvely in Eq.~\eqref{eq:naive_estimation_BR}, observables from $c \to u \,\nu \bar \nu$ transitions represent exceptional tests for the search for NP. In this section, we study the implications of the $SU(2)_L$--link~\eqref{eq:super} in $c \to u \,\nu \bar \nu$ transitions. We use the strongest upper limits on $\mathcal{K}_{L,R}^{N\ell\ell^\prime}$ from high--$p_T$ \cite{Fuentes-Martin:2020lea,Angelescu:2020uug} and charged dilepton data~\cite{Bause:2020auq,Gisbert:2020vjx}, which allow to set constraints on $x_{cu}$. Using Eq.~\eqref{eq:super}, we obtain upper limits on $x_{cu}$ for the different benchmarks {\it i)}-{\it iii)}:
\begin{align}  \label{eq:LU}
x_{cu} &= 3\, r^{\mu \mu} \lesssim 2.6 \,, \quad  (\text{LU}) \\ \label{eq:cLFC}
x_{cu} &= r^{e e}\hspace{-0.1cm}+ r^{\mu \mu }\hspace{-0.1cm}+r^{\tau \tau}  \lesssim 156 \,, \quad (\text{cLFC}) \\ \label{eq:total}
x_{cu} &= r^{ee} \hspace{-0.1cm}+ r^{\mu \mu}\hspace{-0.1cm} +r^{\tau \tau} \hspace{-0.1cm}+2 \,( r^{e \mu}+ r^{e \tau}+r^{ \mu \tau} ) \lesssim 655 \,, \quad (\text{EFT}),
\end{align}
The LU-limit \eqref{eq:LU} is set by dimuon limits which are the most stringent ones. Experimental measurements above the upper limit~\eqref{eq:LU} would indicate a breakdown of LU, while values above~\eqref{eq:cLFC} would imply a violation of cLFC. The quantity $x_{cu}$ enters directly in the branching ratio $\mathcal{B}(c \to u \,\nu \bar \nu)\propto x_{cu}$, as seen in Eq.~\eqref{eq:observable}. Translating the limits \eqref{eq:LU}, \eqref{eq:cLFC} and \eqref{eq:total} in terms of $\mathcal{B}(D^0\to\pi^0\nu\bar{\nu})$,\footnote{Upper limits on other decays modes can be found in Ref.~\cite{Bause:2020xzj}.} we obtain the following upper limits for the three flavor scenarios, 
\begin{align}
    &\mathcal{B}(D^0\to\pi^0\nu\bar{\nu})_{\text{LU}}<5\cdot 10^{-8}~,\label{eq:cuLU}\\
    &\mathcal{B}(D^0\to\pi^0\nu\bar{\nu})_{\text{cLFC}}<2.8\cdot 10^{-6}~,\label{eq:cuCLFC}\\
    &\mathcal{B}(D^0\to\pi^0\nu\bar{\nu})_{\text{EFT}}<1.2\cdot 10^{-5}~.\label{eq:cuEFT}
\end{align}
A branching ratio measurement $\mathcal{B}_{\text{exp}}(D^0\to\pi^0\nu\bar{\nu})$ within $5\cdot 10^{-8} < \mathcal{B}_{\text{exp}}(D^0\to\pi^0\nu\bar{\nu}) < 2.8\cdot 10^{-6}$ would be a clear signal of LU violation. In contrast, a branching ratio above $2.8\cdot 10^{-6}$ would imply a breakdown of cLFC. A very recent search by BES III reported $\mathcal{B}(D^0\to\pi^0\nu\bar{\nu}) < 2.1\cdot10^{-4}$ at 90\% CL \cite{BESIII:2021slf}, which is about one order of magnitude away from our EFT limit~\eqref{eq:cuEFT}.

\subsection{Predictions for beauty}\label{sec:beauty}

In the following we study $b\to s\,\nu\bar\nu$ transitions and their
interplay with $b\,s\,\ell^+\,\ell^-$ and $t\,c\,\ell^+\,\ell^-$ couplings driven by Eq.~\eqref{eq:super}. First, we exploit Eq.~\eqref{eq:super} using the complementarity between $B\to V\,(\text{vector})$ and $B\to P\,(\text{pseudoscalar})$ dineutrino branching ratios, which offers novel tests of LU. Finally, \eqref{eq:super} allows to set improved limits on charged $\tau$ couplings from dineutrino data. 

\begin{figure}[h!]
\begin{minipage}{0.55\linewidth}
A branching ratio measurement outside the red region~\eqref{eq:cone} would clearly signal evidence for LU violation, but if a future measurement is inside this region instead, this may not necessarily imply LU conservation. Outside the light green region the validity of our EFT framework gets broken, see Section~\ref{subsec:EFTtest}. The current experimental $90\,\%$ CL upper limits, $\mathcal{B}(B^0 \to K^{*0}\nu\bar{\nu})_{\text{exp}}\,<\,1.8\,\cdot\, 10^{-5}$ and $\mathcal{B}(B^0 \to K^0\nu\bar{\nu})_{\text{exp}}\,<\,2.6\,\cdot\, 10^{-5}$ from Ref.~\cite{Zyla:2020zbs}, are displayed by hatched bands. The grey band represents the improved EFT limit, $\mathcal{B}(B^0 \to K^0\nu\bar{\nu})_{\text{derived}}\,<\,1.5\,\cdot\, 10^{-5}$, as already briefly mentioned in Section~\ref{subsec:EFTtest}. A measurement between the grey and the hatched area would infer a clear hint of NP not covered by our EFT~\eqref{eq:Heff_noright}. The projected experimental sensitivity ($10\,\%$ at the chosen point) of Belle II with $50\,\text{ab}^{-1}$ is illustrated by the yellow box~\cite{Kou:2018nap}. 
\end{minipage}\hspace*{0.2 cm}\begin{minipage}{0.45\linewidth}
    \centering
    \includegraphics[scale=0.45]{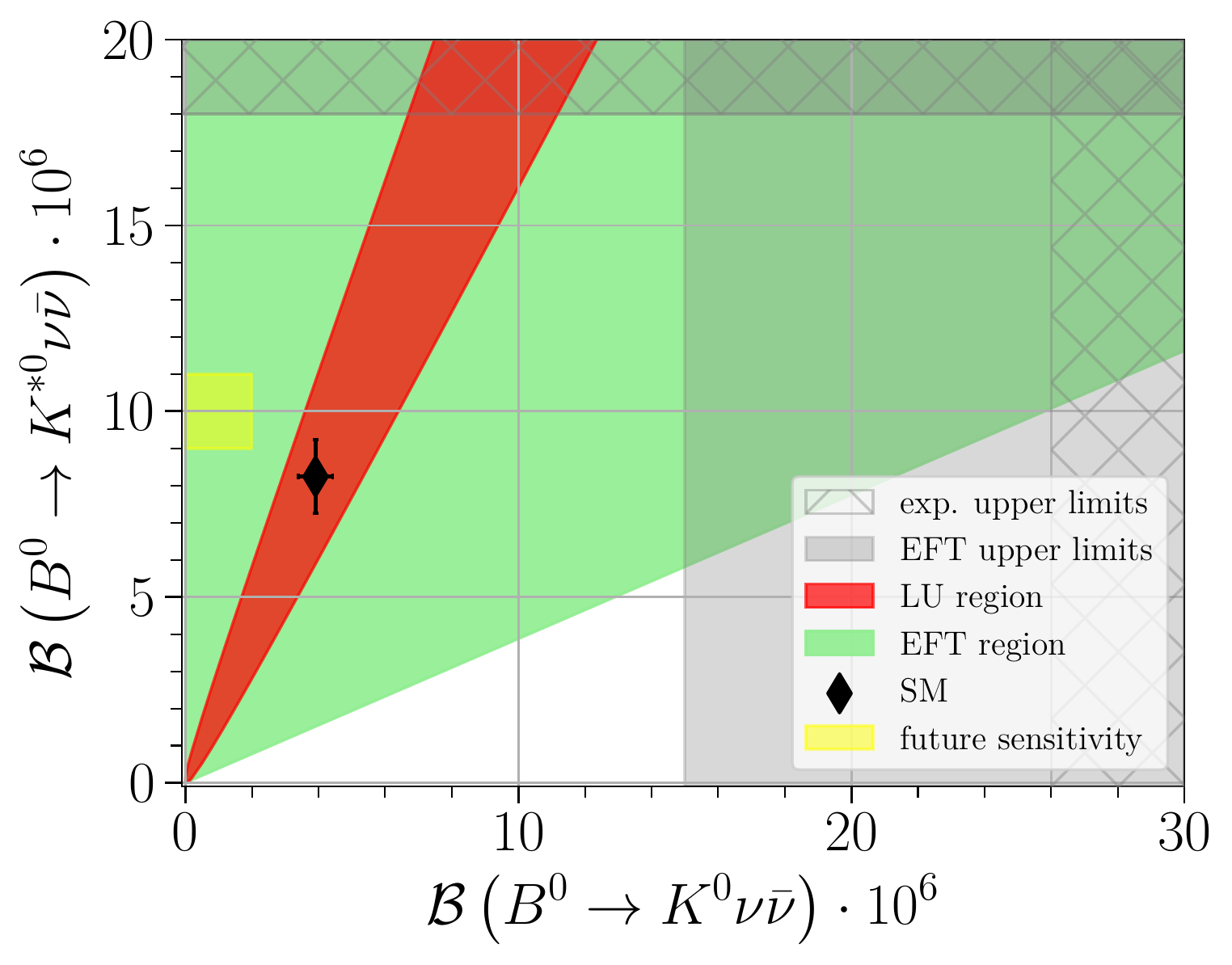}
    \caption{Correlation~\eqref{eq:luregion} between $\mathcal{B}(B^0 \to K^{*0}\nu\bar{\nu})$ and $\mathcal{B}(B^0 \to K^0\nu\bar{\nu})$. Details are given in the main text. }
    \label{fig:plotKstarversusK}
\end{minipage}
\end{figure}

The excellent complementarity between the branching ratios of $B\to V\,\nu\bar\nu$ and $B\to P\,\nu\bar\nu$ decays, $\mathcal{B}(B\to V\,\nu\bar\nu)\,=\,A_+^{B V}\,x_{bs}^+ +\,A_-^{B V}\,x_{bs}^-$ and $\mathcal{B}(B\to P\,\nu\bar\nu)\,=\,A_+^{B P}\,x_{bs}^+$, allows us to write them in the LU limit as~\footnote{The values of $A_\pm^{B V}$ and $A_+^{B P}$ (analogous to $A_\pm^i$ in Eq.~\eqref{eq:observable}) can be found in Ref.~\cite{Bause:2021cna}.}
\begin{align}\label{eq:luregion}
    &\mathcal{B}(B\to V\,\nu\bar\nu)_{\text{LU}}\,=\,\frac{A_+^{BV}}{A_+^{BP}}\,\mathcal{B}(B\to P\,\nu\bar\nu)_{\text{LU}}\,+\,3\,A_{-}^{BV}\,\left|\sqrt{ \frac{\mathcal{B}(B\to P\,\nu\bar\nu)_{\text{LU}}}{3\, A_+^{BP} }}\mp 2\,\mathcal{K}_{R}^{bs\ell\ell}\right|^2~.
\end{align}
The most stringent limits on $\mathcal{K}_{R}^{bs\ell\ell}$ are given for $\ell\ell=\mu\mu$. Performing a 6D global fit of the semileptonic Wilson coefficients $\mathcal{C}_{(7,9,10),\mu}^{(\prime)}$ to the current experimental data on $b\to s\,\mu^+\mu^-$ data (excluding observables polluted by NP effects in electron couplings), leads to $\mathcal{K}_{R}^{bs\ell\ell}=V_{tb}V^{\ast}_{ts}\,(0.37\pm0.28)$.~\footnote{I am grateful to Rigo Bause for providing me with the updated value of $\mathcal{K}_{R}^{bs\ell\ell}$ available in Ref.~\cite{PhDthesis_Bause}.} Figure~\ref{fig:plotKstarversusK} displays the correlation between $\mathcal{B}(B^0 \to K^{*0}\nu\bar{\nu})$ and $\mathcal{B}(B^0 \to K^0\nu\bar{\nu})$, given by Eq.~\eqref{eq:luregion}. The SM predictions $\mathcal{B}(B^0 \to K^{*0}\nu\bar{\nu})_{\text{SM}}=(8.2\,\pm\,1.0)\cdot 10^{-6}$, $\mathcal{B}(B^0\to K^0\nu\bar{\nu})_{\text{SM}}=(3.9\,\pm\,0.5)\cdot 10^{-6}$ are depicted as a black diamond with their $1\sigma$ uncertainties (black bars)~\cite{Bause:2021cna}. 
The values of $\mathcal{K}_{R}^{bs\mu\mu}$, $A_\pm^{B^0 K^{*0}}$, and $A_+^{B^0 K^{0}}$ has been scanned within their $1\sigma$ uncertainties, resulting in the dark red region which represents the LU region, numerically~\cite{Bause:2021cna}
\begin{align}\label{eq:cone}
1.7\lesssim\frac{\mathcal{B}(B^0 \to K^{*0}\nu\bar{\nu})}{\mathcal{B}(B^0 \to K^{0}\nu\bar{\nu})}\lesssim2.6\,.  
\end{align}

The combination of the current experimental 90 \% CL upper limits on $\mathcal{B}(B^0\to K^{*0}\nu\bar{\nu})_\text{exp}<1.8\cdot 10^{-5}$ and $\mathcal{B}(B^+\to K^{+}\nu\bar{\nu})_\text{exp}<1.6\cdot 10^{-5}$ taken from Ref.~\cite{Zyla:2020zbs}, results in the following bounds~\cite{Bause:2021cna}
\begin{align}\label{eq:limits}
    x^+_{bs}\lesssim 2.9~,\quad x^-_{bs}+0.2\,x^-_{bs}\lesssim 2.0~.
\end{align}
Using Eq.~\eqref{eq:limits} together with  Eqs.~\eqref{eq:xpm},~\eqref{eq:super} allows us to set bounds on $\mathcal{K}_L^{tc\ell\ell^\prime}$ and $\mathcal{K}_R^{bs\ell\ell^\prime}$ depending on the lepton flavour assumptions. The limits from
charged lepton data are stronger or similar than the
dineutrino bounds with the exception of $\tau\ell$ couplings,
\begin{align}
    |\mathcal{K}_R^{bs\tau e}|\lesssim 1.4~,\quad|\mathcal{K}_R^{bs\tau\mu}|\lesssim 1.4~,\quad|\mathcal{K}_R^{bs\tau\tau}|\lesssim 1.8~,
\end{align}
which improve the previous ones from Drell-Yan data~\cite{Fuentes-Martin:2020lea,Angelescu:2020uug} by a factor $\sim\,$20~. Constraints on $\mathcal{K}_L^{tc\ell\ell^\prime}$ from dineutrino data are available in Table 6 from Ref.~\cite{Bause:2020auq}, and are stronger than those from collider studies of top plus charged dilepton processes~\cite{CMS:2021nlh,ATLAS:2018avw,CMS:2020lrr}.

\subsection{Predictions for kaons}

Following a similar analysis as in the previous section for $b\to s\nu\bar{\nu}$ transitions, we obtain limits on $\mathcal{K}_L^{cu\ell\ell^\prime}$ and $\mathcal{K}_R^{sd\ell\ell^\prime}$ from the experimental $\mathcal{B}(K^+\to\pi^+\nu\bar{\nu})_{\text{exp}}\lesssim 1.7\cdot10^{-10}$ measurement,~\cite{Bause:2020auq}
\begin{align}\label{eq:kaonlimits}
    -0.019\lesssim\mathcal{K}_L^{cu\ell\ell},\mathcal{K}_R^{sd\ell\ell},\lesssim0.007,\quad \vert\mathcal{K}_L^{cu\ell\ell^\prime}\vert,\vert\mathcal{K}_R^{sd\ell\ell^\prime}\vert\lesssim 0.008~,
\end{align}
with $\ell=e,\mu,\tau$ and $\ell\neq\ell^\prime$. Comparing Eq.~\eqref{eq:kaonlimits} with high-$p_T$ limits~\cite{Fuentes-Martin:2020lea,Angelescu:2020uug}, we observe that dineutrino data provides stronger constraints, by several orders of magnitude.

\section{Conclusions}\label{sec:con}

Dineutrino modes are potential probes for the discovery of physics beyond the Standard Model. In the context of SMEFT, we have shown that $SU(2)_L$-invariance relates dineutrinos $q_\alpha\, q_\beta\,\bar\nu\,\nu$ and charged dilepton couplings $q_\alpha\, q_\beta\,{\ell^\prime}^+\,{\ell}^-$, see Eq.~\eqref{eq:super}. This connection~\eqref{eq:super} provides complementary tests of lepton flavour violation and improved limits on $\tau\,e$, $\tau\,\mu$, $\tau\,\tau$ couplings, exploited for different phenomenological applications including charm, beauty and kaons. Our predictions are well-suited for the experiments Belle II~\cite{Kou:2018nap}, BES III~\cite{Ablikim:2019hff}, and future $e^+ e^-$-colliders, such as an FCC-ee running at the $Z$ mass scale~\cite{Abada:2019lih}.

\section*{Acknowledgements}

I would like to thank Rigo Bause, Marcel Golz and Gudrun Hiller for enjoyable collaborations. Thanks also to the organizers for their effort to make this conference such a successful event. 
This work is supported by the \textit{Bundesministerium f\"ur Bildung und Forschung}--BMBF under project number 05H21PECL2.

\end{document}